\documentclass[twocolumn, amsmath,nobibnotes,nofootinbib, aps, prl, showpacs, 10pt]{revtex4} 
\usepackage[dvips]{graphicx}
\begin{document}
\title{Intragrain electrical inhomogenities and compositional variation of static dielectric constant in LaMn$_{1-x}$Fe$_x$O$_3$}
\author{A. Karmakar$^1$}
\author{S. Majumdar$^1$}
\author{S. Patnaik$^2$}
\author{S. Giri$^1$}
\email{sspsg2@iacs.res.in} 
\affiliation{$^1$Department of Solid State Physics and Center for Advanced Materials, Indian Association for the Cultivation of Science, Jadavpur, Kolkata 700 032, INDIA \\
$^2$School of Physical Sciences, Jawaharlal Nehru University, New Delhi 110067, INDIA}

\begin{abstract}
Bulk dc resistivity and dielectric constant measurements with temperature and frequency have been performed in LaMn$_{1-x}$Fe$_x$O$_3$ (0 $\leq x \leq$ 1.0) as a result of Fe substitution both for the as-synthesized and oxygen annealed samples. Temperature dependence of real part of dielectric constant at different frequencies show a frequency and temperature independent value ($\epsilon_ \textup s$) at low temperature for $x \geq$ 0.15 where $\epsilon_ \textup s$ rises with $x$ reaching a maximum at $x$ = 0.5 and then it decreases sharply at $x$ = 0.70 showing a further increasing trend with the further increase in $x$. The sharp drop of $\epsilon_ \textup s$ at $x$ = 0.70 is correlated with the structural change from rhombohedral to orthorhombic structure. A considerable increase of $\epsilon_ \textup s$ $\sim$ 43 \% is observed for an increase of $x$ from 0.15 to 0.50. Furthermore, $\epsilon_ \textup s$ is increased considerably (up to $\sim$ 17 \% at $x$ = 0.5) due to the oxygen annealing for $x \leq$ 0.50. The low temperature resistivities satisfying Variable Range Hopping model are found to be related with the increase of $\epsilon_ \textup s$ with $x$ for $x \leq$ 0.50. The analysis of the complex impedance and modulus planes at low temperature indicates the electrical inhomogeneities in the grain interior of the compounds.
\end{abstract}


\maketitle

\section{Introduction}
The search for the materials exhibiting large dielectric permittivity is a thrust area of research, because these materials have been recognized as promising candidates for the  applications in electroceramic devices such as capacitors and random access memories. Recent reports on giant dielectric permittivity have directed considerable attention to several new materials such as percolative BaTiO$_3$ -Ni composites \cite{pech},  (Li,Ti)-doped NiO \cite{wu}, epitaxial thin films of Ca$_{\rm 1-x}$La$_{\rm x}$MnO$_3$ \cite{cohn} and nonferroelectric CaCu$_3$Ti$_4$O$_{12}$ \cite{sub,homes}. The origins of large dielectric response in these materials still remain elusive. For example, several possible mechanisms have been proposed to interpret the large dielectric constant in CaCu$_3$Ti$_4$O$_{12}$ such as fluctuations of lattice-distortion-induced dipoles in nanosize domains \cite{homes}, electrode polarization effects due to the different work functions of electrode and sample \cite{luk}, internal barrier layer capacitor (IBLC) effects originating from the insulating grain boundaries surrounding semiconducting grains \cite{sin}, intragranular insulating barrier effects \cite{fu}, etc. Currently, a considerable amount of research is devoted for  understanding the origin of such high dielectric response which triggers special attention for fundamental research as well as for the development of new materials having large dielectric response.
\par
Rare earth manganites with perovskite structure have been recognised as promising candidates for exhibiting multiferroicity \cite{CNR} and colossal magnetocapacitive response \cite{rai} which have been suggested due to the intrinsic phase separation in the grain interior. Recent investigations on the magnetic properties of LaMn$_{1-x}$Fe$_{x}$O$_3$ exhibit the evidence of intrinsic phase separation between ferromagnetic and spin-glass regions in the grain interior for $x$ = 0.30 and 0.50 \cite{patra,thakur,de,liu,bhame}. Although detailed magnetic properties are available for  LaMn$_{1-x}$Fe$_{x}$O$_3$ \cite{kde1,zhou,tong}, the transport mechanism has not been probed explicitly as a result of Fe substitution. It is worthwhile to note that the magnetic phase coexistence  might be the origin of electrical inhomogeneity in the grain interior of the compounds. In this paper, we investigate dc conductivity and dielectric response for different compositions ($x$) and oxygen stoichiometry in LaMn$_{\rm 1-x}$Fe$_{\rm x}$O$_3$. The results indicate the signature of grain interior electrical heterogeneity in the compounds which leads to a considerable change in the magnitude of temperature and frequency independent dielectric constant at low temperature.
 
\section{Experiment}
The polycrystalline compounds LaMn$_{1-x}$Fe$_{x}$O$_3$ with $x$ = 0, 0.15, 0.30, 0.50, 0.70, and 1.0 were prepared by a standard sol-gel technique which is described in our earlier report. \cite{kde1} The final heat treatments were performed in the form of pellets (1.1 cm in diameter and approximately 1 mm in thickness) at 1000$^\circ$C for 12 h in air followed by furnace cooling down to room temperature. A part of each of the samples were further  annealed at 1000$^\circ$C for 12 h in a closed oxygen atmosphere. The single phase of the crystal structure  was confirmed by a BRUKER axs x-ray powder diffractometer (Model no. 8D - ADVANCE). X-ray powder diffraction patterns are shown in figure 1 for the oxygen  annealed samples of all compositions. For $x \leq$ 0.50 the x-ray patterns could be indexed by rhombohedral structure ($R\bar{3}c$) whereas orthorhombic structure ($Pbnm$) is observed for $x \geq$ 0.70 which is consistent with the previous reports \cite{zhou}. In accordance with the previous report \cite{kde1,zhou} the lattice parameters remain almost unaltered in the rhombohedral structure for $x \leq$ 0.30. The maximum intensed lines are highlighted in the inset of figure 1 where one can note that the peak positions remain almost unchanged for $x \leq$ 0.30. At room temperature $^{57}$Fe M\"{o}ssbauer results indicate the absence of precipitation of any iron oxide phase in the compounds. 

\par
Capacitance ($C$) measurements were carried out in the frequency range 20 Hz to 2 MHz using an Agilent E4980A LCR meter which was fitted with a computer for acquiring the data. The measurement was carried out in the temperature range 20 - 300 K using a low temperature cryo-cooler (Advanced Research Systems, USA) wired with coaxial cables. All samples were circular discs in shape and the electrical contacts were fabricated using air drying silver paint. The silver electrodes along with the sample were cured at 150 $^\circ$C for 4 h. The dc resistivity measurements were carried out in a liquid nitrogen cryostat with a trace of helium as an exchange gas. Four probe connection was used for the measurement with air drying silver paint for connecting leads to the sample.
\section{Results and discussions}
\begin{figure}[t]
\vskip 0.0 cm
\centering
\includegraphics[width = 8 cm]{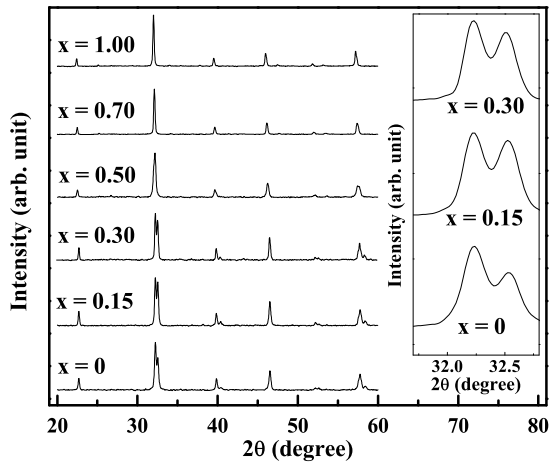}
\caption {X-ray powder diffraction patterns for oxygen annealed samples with $x$ = 0, 0.15, 0.30, 0.50, 0.70 and 1. Inset highlights the maximum intensed peaks.} 
\label{Fig. 1}
\end{figure}

\begin{figure}[t]
\vskip 0.0 cm
\centering
\includegraphics[width = 8 cm]{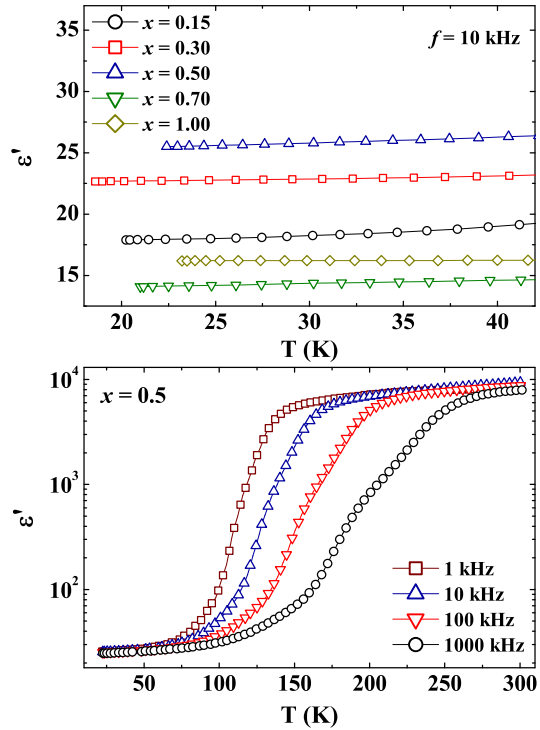}
\caption {Top panel: Temperature dependence of real part of dielectric permittivity ($\epsilon^\prime$) in the low-$T$ range, below $\sim$ 45 K, for oxygen annealed samples with $x$ = 0.15, 0.3, 0.5, 0.7, 1 and $f$ = 10 kHz. Bottom panel: Temperature dependence of $\epsilon^\prime$ for $x$ = 0.5 sample at f = 1 kHz, 10 kHz, 100 kHz and 1000 kHz.} 
\label{Fig. 2}
\end{figure}

\begin{figure}[t]
\vskip 0.0 cm
\centering
\includegraphics[width = 8 cm]{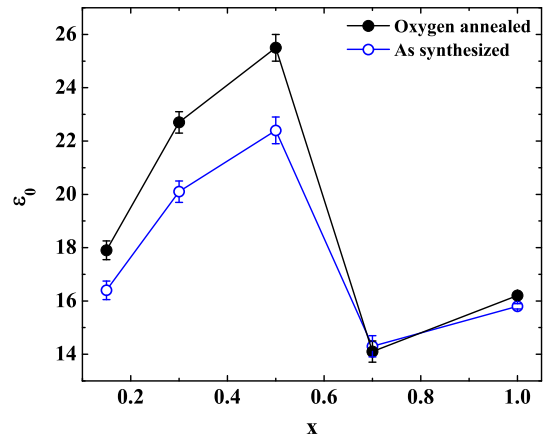}
\caption {Variation of the static dielectric constant ($\epsilon_ \textup s$) with $x$ for both the oxygen annealed and the as-synthesized samples.} 
\label{Fig. 3}
\end{figure}

We have measured the real part of complex dielectric permittivity ($\epsilon^\prime$) as a function of temperature ($T$) and frequency ($f$) for all the compositions. The magnitude of $\epsilon^\prime$ for each sample shows a temperature as well as frequency independent behavior at low temperature which is defined as the static dielectric constant ($\epsilon_ \textup s$). Top panel of figure 2 highlights the temperature variation of $\epsilon^\prime$ at $f$ = 10 kHz up to 45 K exhibiting nearly temperature independent values for oxygen annealed samples with all compositions. The $\epsilon^\prime$ is also independent of frequency in the low temperature region. The bottom panel of figure 2 shows the temperature dependence of $\epsilon^\prime$ measured at $f$ = 1, 10, 100 and 1000 kHz for a representative composition at $x$ = 0.50. The plot exhibits nearly temperature and frequency independent value of $\epsilon^\prime$ below $\sim$ 55 K. The moderately large value of $\epsilon^\prime$ ($\sim$ 10$^4$) is observed at room temperature which is almost frequency independent and it shows weak temperature dependence down to $\sim$ 150 K at $f$ = 1 kHz. Figure 3 shows the variation of $\epsilon_ \textup s$ with $x$. The data for $x$ = 0 is not included in the figure, because temperature and frequency independent region did not develop while cooling down to $\sim$ 20 K up to which our experimental facility is limited. The reported value of $\epsilon_ \textup s$ is found in the range 17.6 - 18 for LaMnO$_3$ \cite{cohn1,neup} which is close to the value of the oxygen annealed sample for $x$ = 0.15. Initially, the value of $\epsilon_ \textup s$ rises with $x$ reaching a maximum at $x$ = 0.5 and then it decreases sharply at $x$ = 0.70 showing a further increasing trend with $x$. We note a considerable increase of $\epsilon_ \textup s$ $\sim$ 43 \% for the increase of $x$ from 0.15 to 0.50. The sharp drop of $\epsilon_ \textup s$ at $x$ = 0.70 is correlated with the structural change from rhombohedral to orthorhombic structure. The filled and the unfilled symbols represent the data for the oxygen annealed and the as-synthesized samples, respectively, exhibiting similar dependence with $x$ where $\epsilon_ \textup s$ is increased further due to the oxygen annealing for $x \leq$ 0.50. At $x$ = 0.50 the value of $\epsilon_ \textup s$ is increased up to 17 \% due to the oxygen annealing. The origin of considerable change of $\epsilon_ \textup s$ with Fe substitution and oxygen annealing will be discussed elsewhere.

\begin{figure}[t]
\vskip 0.0 cm
\centering
\includegraphics[width = 8 cm]{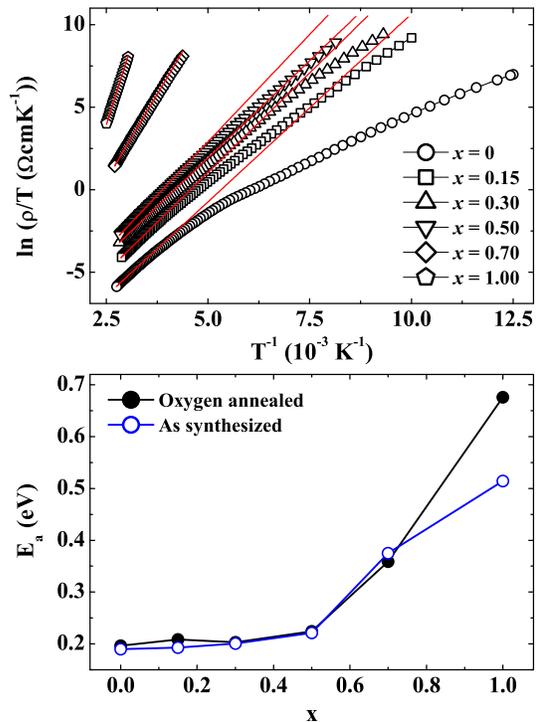}
\caption {Top panel: Temperature dependence of resistivity for the oxygen annealed samples  plotted in the SPH formalism. The solid lines represent the fits according to the SPH model. Bottom panel: Variation of the activation energy ($E_a$), obtained according to the SPH model, with $x$ for both the oxygen annealed and the as-synthesized samples.} 
\label{Fig. 4}
\end{figure}
\begin{figure}[t]
\vskip 0.0 cm
\centering
\includegraphics[width = 8 cm]{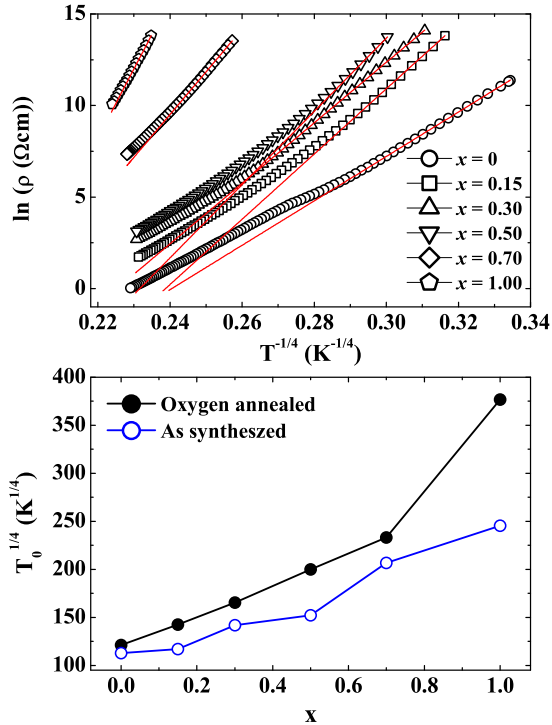}
\caption {Top panel: Temperature dependence of resistivity for the oxygen annealed samples  plotted in the VRH formalism. The solid lines represent the fits according to the VRH model. Bottom panel: Variation of the activation energy ($E_a$), obtained according to the VRH model, with $x$ for both the oxygen annealed and the as-synthesized samples.} 
\label{Fig. 5}
\end{figure}

We measured dc resistivity ($\rho$) for both the oxygen annealed and as-synthesized samples. Semiconducting temperature dependence of resistivity is observed for all the compounds where $\rho$ at room temperature increases monotonically with $x$. The resistivity data are fitted  using Small Polaron Hopping (SPH) as well as Variable Range Hopping (VRH) models.  Figure 4 shows the dc resistivity data fitted to SPH model defined as,
\begin{equation}
\ln\left[\frac{\rho(T)}{T}\right] \propto \frac{E_{\rm a}}{k_{\rm B}T}
\end{equation}
where only limited high temperature region is found to fit satisfactorily. Top panel shows the data corresponding to the oxygen annealed samples. Bottom panel shows the plot of the activation energies ($E_{\rm a}$) with $x$ obtained from the SPH fitting corresponding to both the groups of samples where the values of $E_a$ do not change considerably due to the oxygen annealing for $x \leq$ 0.70 while it differs markedly at $x$ = 1.0. The values are higher for the oxygen annealed sample than the as-synthesized sample. In addition, $E_{\rm a}$ shows a small increasing trend until $x$ = 0.50 above which $E_{\rm a}$ increases sharply. We note that the sharp increase of $E_{\rm a}$ above $x$ = 0.50 is associated with the structural change for $x \geq$ 0.70. 
Top panel of Fig 5 shows the dc resistivity data of the oxygen annealed samples fitted to the VRH model defined as 
\begin{equation}
\ln\left[\frac{\rho(T)}{\rho_0}\right] \propto \left[\frac{T_0}{T}\right]^{1/4}
\end{equation}
which fits satisfactorily in the low temperature region. The constants $T_0$ obtained from the slopes of the plots following VRH mechanism are shown as a function of $x$ for both the groups of the samples in the bottom panel of figure 5. In contrast to the behaviour of resistivity in the high temperature region satisfying SPH mechanism, $T_0$ obtained from VRH fitting is different even at $x$ = 0 due to the oxygen annealing while the difference increases with increasing $x$. Overall temperature dependent resistivity results for all the compositions clearly demonstrate that the effect of Fe substitution and oxygen annealing strongly influences the low temperature resistivity following VRH mechanism and does not influence considerably the high temperature mechanism following SPH model. Increase of oxygen stoichiometry always increases the hopping barrier.  
\begin{figure}[t]
\vskip 0.0 cm
\centering
\includegraphics[width = 8 cm]{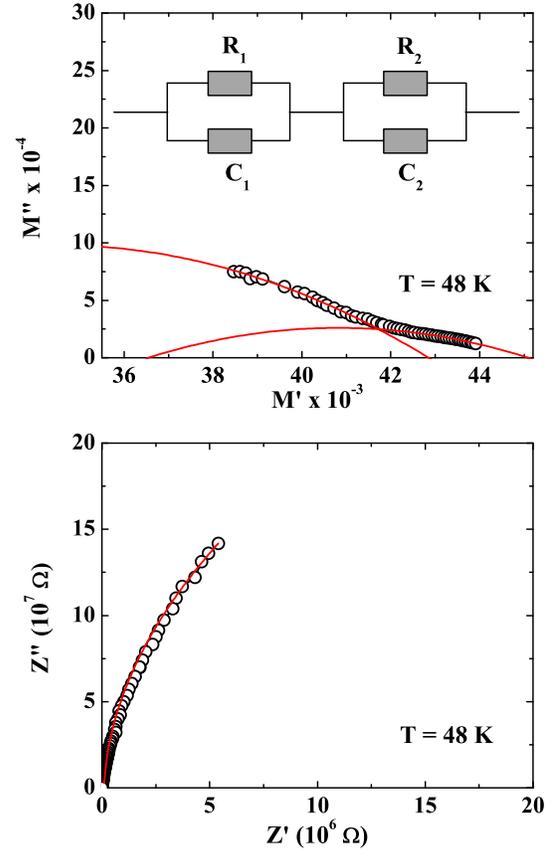}
\caption {Complex modulus (top panel) and complex impedance (bottom panel) plots at 48 K for oxygen annealed sample with $x$ = 0.30. Inset shows the electrical circuit used for the analysis of the plots.} 
\label{Fig. 6}
\end{figure}
\begin{figure}[t]
\vskip 0.0 cm
\centering
\includegraphics[width = 8 cm]{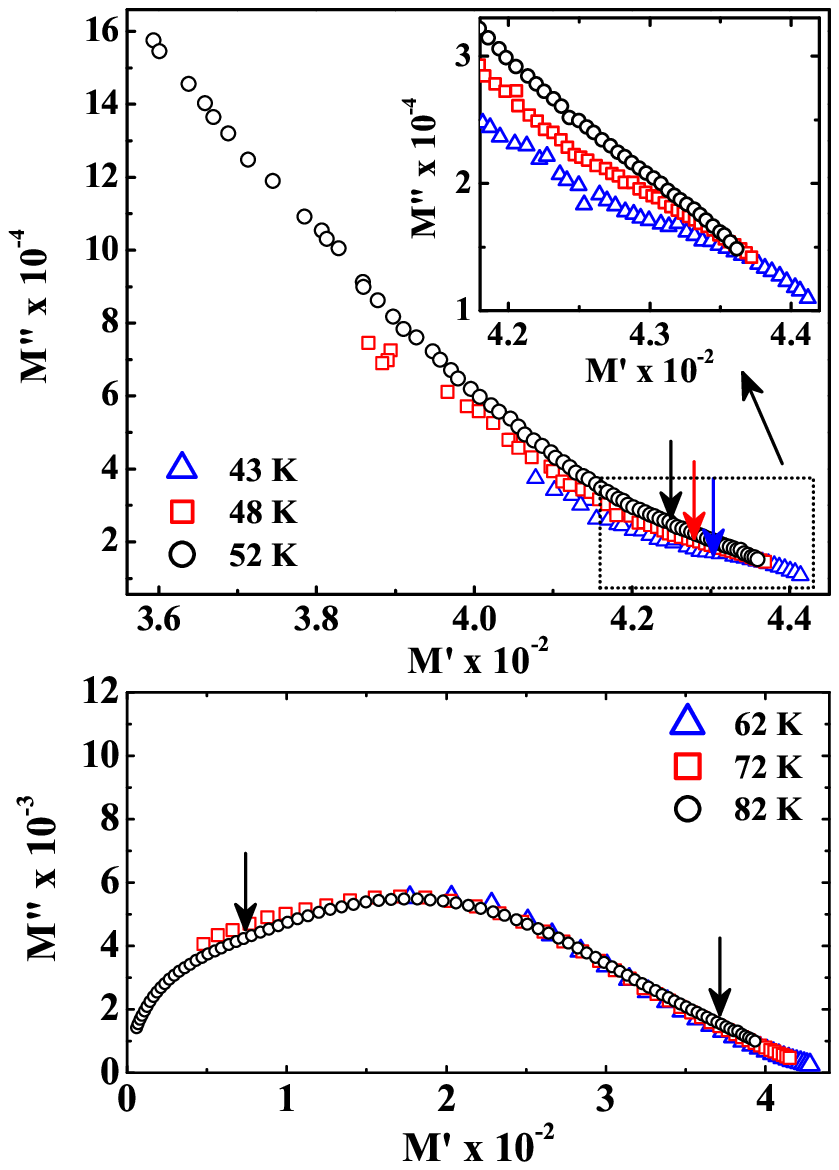}
\caption {Top panel: Complex modulus plot at 43, 48 and 52 K for $x$ = 0.30. Inset zooms the high-$f$ arc. The onset of the arcs are marked by the arrows. Bottom panel: Complex modulus plot at 62, 72 and 82 K for $x$ = 0.30 showing the development of the low-$f$ arc at high-$T$. The crossover among different arcs are marked by the arrows for the plot at 82 K.} 
\label{Fig. 7}
\end{figure}
Recent investigations on the transport properties of hole doped manganites, La$_{1-x}$Ca$_x$MnO$_3$ \cite{cohn1,neup} and La$_{1-x}$Sr$_x$MnO$_3$ \cite{pim} demonstrate the strong correlation between dc conductivity and dielectric response at low temperature where low temperature dielectric constant was found to be correlated with the low temperature hopping mechanism of dc conductivity satisfying VRH process \cite{neup}. In the present investigation a strong correlation between low temperature conductivity and dielectric constant is noticed where VRH hopping barrier increases associated with an increase in $\epsilon_ \textup s$ while $x$ is increasing in the range 0.15 $\leq x \leq$ 0.50. A sharp drop in $\epsilon_ \textup s$ is observed at $x$ = 0.70 where a structural change is noticed which is consistent with the results observed in the low hole doped La$_{1-x}$Ca$_x$MnO$_3$ \cite{neup}. At low temperature the electrode polarization effect, due to different work functions of the electrode and sample, is negligible when the dielectric response is independent of frequency and temperature. The values of $\epsilon_ \textup s$ as a function of $x$ in figure 3 is thus attributed to the grain interior mechanism or grain interior and grain boundary mechanism.
\par
In order to the identify the origin, the complex impedance and modulus plane plots have been used at temperatures where sample-electrode contribution to the response is negligible. The complex-$M$ and complex-$Z$ plots at 48 K are shown in figure 6 for oxygen annealed sample with $x$ = 0.30 where dielectric constant is almost independent of frequency and temperature. The $Z$-plot indicates a portion of a single semicircle (bottom panel of figure  6) whereas the $M$-plot (top panel of figure 6) clearly resolves two semicircles. Top panel of figure 7 shows two other $M$-plots in the same temperature regime at 43 and 52 K together with the plot at 48 K. The onset of the arcs at high-$f$ are marked by the arrows which are highlighted in the inset. In order to identify the microstructural contributions to the semicircles in the complex modulus plots we examine similar plots at little higher temperatures at T = 62, 72 and 82 K which are shown in the bottom panel of figure 7. One can note that as the temperature is increased a third arch gradually develops at the low frequency side. The arch is fully developed in the plot at 82 K where the signature of crossover among  different arcs are marked by the arrows. With the further increase in temperature the high-$f$ arc gradually shifts out of the frequency window.
\par
For an interpretation of the experimental data, it is convenient to have a model equivalent circuit that provides a realistic representation of the electrical properties. The simplest equivalent circuit with a serial array of parallel $RC$ elements is used to interpret the complex plane plot where modulus plot gives emphasis to those elements with smaller  capacitances and impedance plot highlights those with largest resistance \cite{mac}. Moreover, it's well known that modulus formalism has the advantage of eliminating the contact contributions. This plot resolves semicircles when $C$-values are comparable in magnitude \cite{hodge1, hodge2}. Thus we analyze the plot using two segments of parallel $R$ and $C$ elements in series (inset of figure 6). The capacitance values are calculated from the intercepts of the arcs on the real axis of the complex-$M$ plot using the formula $C$ = $\epsilon_0$/$M_{int}$, where $C$ is the estimated capacitance, $\epsilon_0$ is the free space permittivity and $M_{int}$ is the intercept on the real-$M$ axis \cite{Sinclair}. The values are $C_1$ = 0.53 nF and $C_2$ = 1.03 nF which are comparable to each other. Typical capacitive value of the sample-electrode interface is $\sim$ 10$^{-7}$ F. Here small values of $C_1$ and $C_2$ are attributed to the grain interior microstructure where  appearance of a new low-$f$ arc at little higher temperatures for $T \geq$ 62 (bottom panel of figure 7) is suggested due to the grain boundary effect. The corresponding values of $R_1$ and $R_2$ must be calculated similarly from the intercepts in the complex impedance ($Z$) plot where the single arch represents the component with resistance much larger than the other component \cite{Sinclair}. Here the complex-$Z$ plot shows only a portion of a single arc at the high-$f$ region within our available frequency window where, $R$ = 4.2 M$\Omega$. Since the high-$f$ arc generally corresponds to the intrinsic contribution of the sample, high $R$-value obtained from the analysis gives us the high resistive intrinsic component at 48 K for $x$ = 0.3 in LaMn$_{1-x}$Fe$_x$O$_3$. The analysis of the complex-$M$ and $Z$ plane plots indicates that the two high-$f$ arcs in complex-$M$ plot correspond to two different intragrain regions having comparable capacitance values but with largely different conductivities. 

We note that the values of resistivity at 300 K are 2.2 and 3.2 $\times$ 10$^6$ $\Omega$cm for LaMnO$_3$ and LaFeO$_3$, respectively where the substitution of Fe in LaMn$_{1-x}$Fe$_x$O$_3$ significantly increases the resistivity as well as the VRH hopping barrier at low temperature. The considerable increase of $\epsilon_ \textup s$ associated with the increase of VRH hopping barrier with increasing Fe substitution ($x \leq$ 0.50) is suggested to be a result of the intragranular electrical inhomogeneity between semiconducting and insulating conductivities where Fe substitution increases the insulating phase resulting in the increase of $\epsilon_ \textup s$. Moreover, oxygen annealing further increases the insulating phase leading to a further rise of VRH hopping barrier and $\epsilon_ \textup s$ as compared to the as-synthesized samples. Recent reports on the magnetic properties of the sample with $x$ = 0.30 in LaMn$_{1-x}$Fe$_x$O$_3$ clearly demonstrate the grain interior phase separation between ferromagnetic and spin-glass phases \cite{patra,thakur,de,liu,bhame}. The magnetodielectric measurements on the sample $x$ = 0.30 do not show any signature of change in the dielectric constant due to the application of external magnetic field up to 30 kOe. Nevertheless,  it is understood that the conductivities of the ferromagnetic phases are much higher than that of the spin-glass phases which might be correlated with the electrical inhomogeneities in the grain interior of the compounds.

\section{Summary}
The low temperature static dielectric constant ($\epsilon_ \textup s$) is increased considerably ($\sim$ 43 \%) with  the Fe substitution in LaMn$_{1-x}$Fe$_x$O$_3$ for $x \leq$ 0.50. The value is further increased due to oxygen annealing for $x \leq$ 0.50 where $\epsilon_ \textup s$ is increased up to $\sim$ 17 \% at $x$ = 0.50. A sharp change of $\epsilon_ \textup s$ is observed for $x >$ 0.50 which is associated with the structural change. The low temperature dc resistivities following VRH mechanism are found to be correlated with the increase of $\epsilon_ \textup s$ as a result of Fe substitution for $x \leq$ 0.50 and due to the effect of oxygen annealing. A simplified equivalent circuit with a serial array of parallel $RC$ elements is used to interpret the plots of complex impedance and modulus planes which suggest the existence of intragranular electrical inhomogeneities in the compounds. 

\noindent
{\bf Acknowledgement}
S.G. wishes to thank Department of Science and Technology, India (Project No. SR/S2/CMP-46/2003) for the financial support.

\end{document}